# ARGON ATOMISTIC SIMULATIONS WITH EM3, A NEW MOLECULAR MECHANICS PROGRAM

Andrew Rohskopf

Massachusetts Institute of Technology
May 2018

# ACKNOWLEDGEMENTS

EM3 was developed in part of the Fundamentals of Nanoengineering Course at MIT taught by Professor Nicolas Hadjiconstantinou, who offered valuable lectures on molecular dynamics methods. Help was also offered by the TA Mojtaba Oozroody, who patiently provided advice and answered questions.

# TABLE OF CONTENTS

# LIST OF TABLES

# LIST OF FIGURES

x

# LIST OF SYMBOLS AND ABBREVIATIONS

| | |
|---|---|
| c-Ar | Crystalline argon |
| EM3 | Easy $M_1$odular $M_2$olecular $M_3$echanics |
| $L_\alpha$ | Length of simulation box in $\alpha$ Cartesian direction |
| LJ | Lennard-Jones |
| KE | Kinetic energy |
| MC | Monte Carlo |
| MD | Molecular dynamics |
| MSD | Mean square distance |
| $n$ | Number density (same as mass density in LJ units) |
| N | Number of atoms |
| P | Pressure |
| PE | Potential energy |
| T | Temperature |
| $\tau$ | Lennard-Jones time unit |
| $\alpha, \beta, \gamma$ | General Cartesian coordinate $x, y, z$ |
| $\varepsilon$ | Lennard-Jones energy parameter |
| $i, j, k$ | General indices in atomistic summations |
| $\sigma$ | Lennard-Jones length parameter |

# SUMMARY

Argon molecular dynamics (MD) simulations are performed with a newly developed MD program, Easy M$_1$odular M$_2$olecular M$_3$echanics (EM3). The program was developed in an object-oriented fashion containing classes for each critical part of a functioning MD program. An organizational scheme for a general molecular mechanics program is therefore presented, along with the framework of the EM3 program. With the modular nature and open-source availability, EM3 can serve as a learning tool for newcomers to molecular simulations and code organization via object-oriented programming. Validations of the code are presented in comparison with Monte Carlo (MC) simulations of liquid argon at different densities and temperatures. A calculation of the self-diffusion coefficient for liquid argon is also performed, exhibiting the extendibility of EM3. This report comes packaged with the EM3 source code and examples in the `/src` directory and `/examples` directory, respectively; all code and inputs are available for review when following the explanation of algorithms and examples.

# CHAPTER 1.     EASY M$_1$ODULAR M$_2$OLECULAR M$_3$ECHANICS (EM3)

## 1.1   Motivation

A plethora of powerful open-source atomistic simulation programs are available to the public, including LAMMPS[1], GROMACS[2], and many others[3-5]. With the current availability of powerful atomistic simulation programs, the development of another might seem futile. EM3, however, possesses a goal that separates it from the rest of the currently available programs. Unlike existing programs possessing goals centered around simulation performance, the goal of EM3 is more educational and simplistic. The educational goal centers around the simple (easy) objective-oriented (modular) C++ source code of EM3, which aside from being powerful and efficient, seeks to bridge the gap between less powerful but easily understood MATLAB and Python codes scattered about the internet and more powerful but complicated larger programs like LAMMPS and GROMACS (also written in C++ and C, respectively). This educational goal is readily realized by considering the seemingly lack of intensive developer guides for existing programs; this issue may hinder a new researcher, such as a student beginning work in atomistic simulations, from transcending the role of a simple user to a developer, thus gaining the ability to perform more meaningful work. A commonality across the commonly used MD programs is that they are object-oriented; this is a necessary component when developing extendable open-source programs that can readily be modified by any user. Newcomers to the field may find that there is much to learn in terms of object-oriented programming and its relation to organizing a proper MD code. This is where EM3 comes in; it provides a framework for modularly organized atomistic simulation codes, and its open-source availability allows beginners to familiarize themselves with the type of application of object-oriented C++ that is seen in larger and more powerful programs such as LAMMPS. EM3 strives to be a "light" (easy) version of the larger MD programs like LAMMPS and GROMACS, without sacrificing computational power since it is written in the powerful C++ language and supports parallel compilation with MPI. While parallelized MD is not yet implemented, the EM3 source code includes MPI headers and initializers and is compiled using MPI, so that future work can result in parallelization.

The main source file for EM3, `main.cpp`, begins an instance of an EM3 class, defined in `em3.cpp`, which is the driver of the MD simulation. EM3 initializes instances of the various classes existing in the program, and then performs the actual MD simulation. These classes are critical to the organizational modularity of EM3, and many other programs, so a discussion is warranted for each class, or module. ***This report comes packaged with the EM3 source code in the `/src` directory, so all code is available for review when following the explanation of algorithms and schemes.***

**1.2  Input Class & Files**

A class for gather user-input is critical for generally performing a MD simulation using arbitrary settings, and without editing code each time a setting is changed. The Input class, defined in input.cpp, reads settings associated with a MD simulation from a file called INPUT, along with data describing the system being simulated via a file called CONFIG. Simply executing the EM3 program in the same directory as these files will result in the inputs being read. The current INPUT file is structured as follows:

```
THIS LINE IS IGNORED
    nsteps: 1100000
    nequil: 100000
    nout: 100
    max_neigh: 200
    rc: 3
    neighcount: 0
    newton: 0
    offset: 0
    mass: 6.6335209e-26
    temperature: 158
    timestep: 1
    epsilon: 1.65e-21
    sigma: 3.4e-10
```

**Figure 1 – INPUT file for the EM3 program. It is important that this format is exactly followed, including the order in which lines are read, since no error catching is currently implemented in the program.**

The first line of the file is ignored. The next line shown by `nstep` declares the number of timesteps to perform a MD simulation. A simple static calculation can be performed by setting this value to zero. The `nequil` line determines how many timesteps an equilibration run will be performed under constant temperature. The total simulation time is therefore the sum of `nstep` and `nequil`. The next line, `nout`, tells EM3 to output data in increments of this value. The maximum number of neighbors are declared on the next line by declares are the number of timesteps, number of timesteps to skip before outputting data, the maximum number of allowable neighbors for each atom. A smaller number means that a smaller neighbor-list array is initialized, thus requiring less memory. The cutoff (in LJ units) is declared in the next line by `rc`. The neighbor-list counting is declared by `neighcount` and can either be 1 for full neighbor-list calculation (i.e., looping over all neighbors $j \neq i$) or 0 for a half neighbor-list calculation (i.e., looping over all neighbors $j > i$). The `newton` keyword turns on (set to 0) or off (set to 1) the application of Newton's third law when calculating forces between atoms. For the applications in this report, using a half neighbor-list, Newton's third law should be applied to balance the forces in pair interactions. The next contains the `offset` keyword, which can be set to 0 to subtract the potential energy offset due to a cutoff, or 1 to exclude the offset correction. The next settings are the atomic mass in kg, temperature in K, timestep in fs, Lennard-Jones (LJ) parameter ε in J units, and LJ parameter σ in m units.

While these declarations in the INPUT file involve settings about the MD simulation, the CONFIG file is used to declare structural settings about the system of atoms under consideration. The format of the CONFIG file is illustrated below.

```
natoms ntypes
box_xx box_yy box_zz
type id x y z
type id x y z
type id x y z
type id x y z
```

**Figure 2 – Format of CONFIG file. In this case, there are 4 lines declaring atom types, IDs, and positions and therefore the `natoms` keyword should be equal to 4.**

The first line of CONFIGS declares integers `natoms` and `ntypes` which are the number of atoms and number of atom types (unique species) in the system, respectively. The next line declares cubic box dimensions in all three Cartesian directions. The following lines (which there should be `natoms` of) declare the type, ID number, and Cartesian position of each atom. Quantities declared in the CONFIG file must be in reduced LJ units.

The input class also contains a function `initialize()` which converts necessary units to LJ units and initializes the velocities according to the desired temperature. This done for each velocity component by sampling a normally distributed number on the interval $[-1,1]$, and then multiplying the number by $\sqrt{T}$ in LJ units. The center of mass velocities are also subtracted from the initial velocities so that the simulation begins with zero linear momentum (otherwise the system will translate). The center of mass velocity for each Cartesian component $\alpha$ is calculated via $v_{com}^{\alpha} = \sum_i v_i^{\alpha}$ (since mass is unity in LJ units) for all atoms $i$ and then subtracted from each initial velocity component for every atom.

### 1.3    Neighbor Class

The neighbor class (neighbor.cpp) uses the current atom positions and user-defined cutoff to generate a neighbor-list. In EM3, the current neighbor-list can be referenced through any class using the standard C++ arrow pointer `neighbor->neighlist`, and it is a 2D array. The first dimension of the neighbor-list in EM3 are the atoms $i$ existing in the system, while the second dimension contains the indices of neighbors $j$ of atom $i$. The neighbor-list is generated according to the minimum image convention, which allows the original atoms to move outside the simulation box without enforcing the coordinates to reside in the box. This is achieved, for every Cartesian coordinate $\alpha = (x, y, z)$, by finding the Cartesian displacement coordinates $\alpha_{ij} = \alpha_i - \alpha_j$ between all neighbors, and then applying the minimum image convection via

$$\alpha_{ij} = (\alpha_i - \alpha_j) - L_\alpha \text{round}\left[\frac{1}{L_\alpha}(\alpha_i - \alpha_j)\right] \qquad (1)$$

for every Cartesian direction $\alpha$, the round function rounds the argument to the nearest integer, and $L_\alpha$ is the length of the simulation box in the $\alpha$ direction. While not inherently intuitive, this transformation allows to use the un-altered coordinates of original atoms in the simulation box by letting them traverse through all space, in which they interact with the nearest images of other atoms in the simulation. In EM3, the positions of every atom, including minimum image neighbors for each atom $i$ and their positions defined by Equation 1, are stored in the 2D array `x`. The first dimension of `x` is the index of atom $i$ and the 2$^{nd}$ dimension is the Cartesian coordinate indexed by `0, 1` or `2`. For example, the y-coordinate of atom 36 is found via `x[35][1]` (note that indices start from 0 in C++). The current positions in the simulation are accessed in any class via the pointer `neighbor->x`. Current atomic positions are therefore stored by the neighbor class and referenced by other classes to do calculations at a timestep. There may be a better scheme for storing current atomic positions in a class other than then neighbor class but optimizing this organization can be a part of future work.

The size of the `neighbor->neighlist` array is `natoms` (number of atoms declared in the CONFIG file) in the first dimension, and `neighmax` in the 2$^{nd}$ dimension. The first dimension therefore indexes the original atoms $i$ that were declared in the CONFIG file, from 0 to `natoms-1` in C++. The 2$^{nd}$ dimension contains integers (indices) $j$ that are neighbors of $i$. These indices can run from 0 to `(natoms*neighmax)-1`, since any of the original atoms or any of their neighbors can be a possible neighbor. It is important to distinguish between indexing the original atoms declared in the CONFIG file and minimum images of these atoms according to Equation 1. Original atoms contain the indices 0 to `natoms-1` but *images* of these atoms can contain different indices. Therefore, an if-statement inside neighbor.cpp determines whether a neighbor $j$ of atom $i$ is one of the originally declared atoms in CONFIG. If it is, the original index is stored in `neighlist`. Otherwise, a new index is assigned to this atom, which is an image of the original atom, and this new index is stored in `neighlist`. This results in a scheme where the indices stored in the `neighlist` array always correspond to their atom positions in the `neighbor->x` array. This is useful for the potential class, when looping over atoms to calculate contribution to potential energy, since all that is required is the `neighlist` array to obtain positions of these neighbors. For example, when looping over neighbors of atom $i$, we may find that the contents of `neighlist[i]` are `2, 3,` and `198`. This means that we can obtain the positions of these neighbors by retrieving `x[2], x[3],` and `x[198]`. The usefulness of this scheme is realized by considering, for example, if atom index `198` is larger than `natoms-1`. This is entirely possible since the atom indexed by `198` may be an *image* of another atom within the originally declared box in CONFIG. We still want to keep track of this image and its position since it is a neighbor of atom $i$, and this will be useful when calculating the potential energy and forces, which depend on the positions of all atoms and their neighbors.

## 1.4 Potential Class

The potential class (potential.cpp) codes the potential and calculates forces, potential energy and pressure. The neighbor class is called before the potential class, so the potential class has access to the neighbor-list of atomic positions and their neighbors. A double loop over all neighbors is necessary for a simple 2-body pair potential calculation, such as the Lennard-Jones (LJ) potential used in Chapter 2. The modularity of EM3 allows this class to be replaced with any other class encoding any other potential, but for this report the LJ potential in non-dimensional units or "LJ units" is used. LJ units are defined in Table 1.

**Table 1 – Reduced LJ units (denoted by asterisks *) used in this report.**

|  | Conversion |
|---|---|
| Length, $x^*$ | $x^* = \dfrac{x}{\sigma}$ |
| Energy, $E^*$ | $E^* = \dfrac{E}{\varepsilon}$ |
| Mass, $m^*$ | $m^* = \dfrac{m}{m} = 1$ |
| Time, $t^*$ | $t^* = t\sigma\sqrt{\dfrac{m}{\varepsilon}}$ |
| Force, $f^*$ | $f^* = f\dfrac{\varepsilon}{\sigma}$ |
| Temperature, $T^*$ | $T^* = T\dfrac{k_B}{\varepsilon}$ |
| Velocity, $v^*$ | $v^* = v\dfrac{t^*}{\sigma} = v\sqrt{\dfrac{m}{\varepsilon}}$ |

The LJ potential energy $U$ in terms of reduced units is given by Equation 2

$$U = \sum_i \sum_{j>i} 4\left(\frac{1}{r_{ij}^{12}} - \frac{1}{r_{ij}^{6}}\right) \tag{2}$$

where $i$ runs over all atoms and $j$ is indexed to be greater than $i$ to avoid double counting. In reduced units, the potential energy in Equation 2 depends only on the reduced interatomic distance $r_{ij}$. Taking the negative gradient of $U$ in a Cartesian direction $\alpha$, we obtain the force $f_i^\alpha$ on atom $i$ in the $\alpha$ direction via Equation 3.

$$f_i^\alpha = -\frac{\partial U}{\partial \alpha_i} = \sum_{j>i} 48\left(\frac{1}{r_{ij}^{14}} - \frac{1}{2}\frac{1}{r_{ij}^{8}}\right)\alpha_{ij} \tag{3}$$

The forces can be used to calculate configurational (static) contributions to the stress tensor $P_{\alpha\beta}$ for Cartesian directions $\alpha$ and $\beta$ via

$$P_{\alpha\beta} = \frac{1}{V}\sum_i \sum_{j>i} \alpha_{ij} f_{ij}^\beta \tag{4}$$

where $V$ is the system volume and $f_{ij}^\beta$ is the force on atom $i$ due to atom $j$ in the $\beta$ direction. Pressure in reduced LJ units comes naturally by substituting reduced Cartesian coordinates and forces into Equation 4, which yields the same result except now $V$ is the reduced volume. The loops that calculate the potential energy via Equation 2, forces via Equation 3, and configurational pressure via Equation 4 are in the potential class. After this calculation the potential class stores the potential energy and forces as public variables, where they can be accessed from any other class in EM3 via pointers `potential->pe` and `potential->f`, respectively.

The potential and forces of Equation 2 and Equation 3 are calculated in EM3 by looping over all atoms $i$ and their neighbors $j$, along with the pressure in Equation 4. This is easily done by referencing the neighbor list generated by the neighbor class (i.e., `neighbor->neighlist`) as well as the number of neighbors for each atom $i$ also generated by the neighbor class (i.e., `neighbor->numneigh`). The potential loop over neighbors $j$ therefore contains `numneigh[i]` loops for atom $i$. Further details of this implementation can be seen in the potential.cpp file, along with the coded potential in Equation 2, forces in Equation 3, and configurational pressure in Equation 4.

## 1.5 Update Class

The update class (update.cpp) utilizes the forces from the potential class to obtain the accelerations $a_i^\alpha$ in the $\alpha$ Cartesian direction on all atoms $i$ via Newton's 2nd law, $f_i^\alpha = m_i a_i^\alpha$. Since LJ units for a monoatomic system has $m_i = 1$ for all atoms, according to Table 1, the accelerations for a configuration are simply the forces associated with that configuration (i.e., $a_i^\alpha = f_i^\alpha$). Once the accelerations are known, the positions and velocities at the next timestep can be determined via the velocity Verlet algorithm. The Verlet algorithm as it is coded in the EM3 update class is shown in Figure 3. Note the convenient used of arrow pointers when prompting the generation of a neighbor-list using the neighbor class via `neighbor->generate()`, from which the potential class is used to calculate the potential energy and forces using the new neighbor-list via

potential->calculate(). The accelerations are simply the forces in LJ units, which explains the assignment of new accelerations via potential->f.

```
for (int i=0; i<natoms; i++){

    v[i][0] += 0.5*dt*a[i][0];
    v[i][1] += 0.5*dt*a[i][1];
    v[i][2] += 0.5*dt*a[i][2];
    x[i][0] += dt*v[i][0];
    x[i][1] += dt*v[i][1];
    x[i][2] += dt*v[i][2];
}

// Update neighborlist

neighbor->generate();

potential->calculate();

a = potential->f;

// Perform final step of Verlet algorithm

for (int i=0; i<natoms; i++){
    v[i][0] += 0.5*dt*a[i][0];
    v[i][1] += 0.5*dt*a[i][1];
    v[i][2] += 0.5*dt*a[i][2];
}
```

**Figure 3 – C++ Verlet integration used in the EM3 update class.**

First the velocities for each atom and Cartesian direction are updated at a half timestep $\frac{1}{2}\delta_t$ via

$$v_i^\alpha\left(t + \frac{1}{2}\delta_t\right) = v_i^\alpha(t) + \frac{1}{2}\delta_t a_i^\alpha \qquad (5)$$

followed by an update of positions to the next full timestep via Equation 6.

$$\alpha_i(t + \delta_t) = \delta_t v_i^\alpha \left(t + \frac{1}{2}\delta_t\right) \tag{6}$$

The velocities at the next timestep are given by

$$v_i^\alpha(t + \delta_t) = \frac{\delta_t}{2} v_i^\alpha \left(t + \frac{1}{2}\delta_t\right) \alpha_i(t + \delta_t) \tag{7}$$

which completes the timestep. Equations 5-7 are iterated through timesteps, and timesteps in EM3 are driven in the EM3 class (em3.cpp) by simply calling `update->integrate()`, which encodes the algorithm in Figure 3.

Verlet integration by itself conserves total energy if the potential yields conservative forces and geometry is defined in geometrically invariant manner (i.e., translational and rotational invariance are conserved via the interatomic distance geometric descriptor in the LJ potential). This yields sampling of phase space in the statistical mechanical microcanonical (NVE) ensemble with constant number of particles $N$, constant volume $V$, and constant energy $E$. The temperature is therefore allowed to fluctuate. It is often of interest, however, to calculate system properties at a specific temperature, which we will do in Chapter 2 of this report. To do so, a thermostat must be applied to the system. EM3 currently only uses the Andersen thermostat, which works by rescaling the velocities after Verlet integration via

$$v_i^\alpha = v_i^\alpha \sqrt{\frac{T_0}{T}} \tag{8}$$

where $T_0$ is the desired temperature and $T$ is the actual temperature. Rescaling every velocity in this manner will yield a temperature that is held at the desired temperature but modifying the equations of motion result in a lack of energy conservation. The resulting dynamics samples phase space consistent with a canonical NVT ensemble. EM3 performs this rescaling every timestep. Although a change could easily be made to perform the rescaling a defined number of timesteps to improve computational expense, it was not deemed necessary at this time. With the core MD algorithm now established, it useful to understand how various system quantities are computed in the EM3 program, including the temperature that is required for Equation 8.

### 1.6 Compute Class

The compute class (compute.cpp) can be called by any other class in EM3 to compute various system quantities. This provides an organization framework to store functions that can be called anywhere else in the modular code to compute system quantities such as temperature and total pressure in any class. The system temperature, kinetic energy ($KE$), and total energy are computed by pointing to and calling the function `compute->compute_ke()`. This function first calculates the kinetic energy

$$KE = \frac{1}{2}\sum_i m_i v_i^2 \tag{9}$$

where $m_i = 1$ for a monoatomic system in LJ units. The temperature is then given by

$$T = \frac{2}{3}\frac{(KE)}{Nk_B} \tag{10}$$

where $k_B = 1$ in LJ units. The total energy $E$ is simply $E = U + KE$. Given the temperature, the total pressure $P$ can now be given as the sum of kinetic and static contributions via

$$PV = Nk_BT + \frac{1}{3}(P_{xx} + P_{yy} + P_{zz}) \tag{11}$$

where $P_{xx}$, $P_{yy}$, and $P_{zz}$ are diagonal stress tensor components calculated in the potential class using Equation 4. Chapter 2 of this report includes a diffusion coefficient calculation for liquid argon, and therefore another function was added to the compute class to compute the mean squared displacement (MSD) $\langle \Delta r^2(t) \rangle$ of atoms from their initial positions at a given time. MSD is computed by calling the function `compute->compute_msd()`, which uses the equation

$$\langle \Delta r^2(t) \rangle = \frac{1}{N} \sum_i^N \| r(t) - r(0) \|^2 \tag{12}$$

where $r(t)$ is the atomic position vector at time $t$, $r(0)$ is the atomic position vector at the beginning of the simulation, and $\| r(t) - r(0) \|$ is the norm of the difference between these two vectors. The MSD in Equation 12 can be used with the Einstein relation[6,7], which relates $\langle r^2(t) \rangle$, the mean square distance over which particles have moved in time $t$, to the diffusion coefficient $D$ according to Equation 13[7]

$$\frac{\partial \langle r^2(t) \rangle}{\partial t} = 2dD \approx \frac{\partial \langle \Delta r^2(t) \rangle}{\partial t} \tag{13}$$

where $d$ is the dimensionality of the system ($d = 3$ for a 3D system). The Einstein relation allows us to plot $\langle \Delta r^2(t) \rangle$ as a function of time and then determine the diffusion coefficient as one sixth the slope of the plot for a 3D system. The results of this procedure for argon will be shown in Chapter 2.

### 1.7 Output, Memory & Timer Classes

While all the aforementioned classes perform the important parts of a MD simulation or molecular statics calculation, the output class (output.cpp), memory class (memory.cpp) and timer class (timer.cpp) function as convenient utilities more than anything else. The output class encodes functions that can write out quantities such as positions during a simulation for viewing. The memory class is essential in initializing memory for all the arrays used in EM3, such as the forces, positions and neighbor-list, as well as freeing the memory when the program is complete. The timer class simply keeps track of how long a job has been running and is useful for gauging program performance. The memory and timer class are convenient utilities for allocating and deallocating arrays and keeping track of program runtime; both classes are based on the corresponding memory and timer classes of the ALAMODE program by Terumasa Tadano[8], but these classes have no contribution or effect on MD simulations performed by EM3 – they are simple utilities for program memory and time management. Not much else is to be said about these classes, as they are simply utilities that bring convenience to the modular nature of the code.

### 1.8 Compiling, Running & Examples

The generality of EM3, written in C++ without the use of external libraries, results in a simple compilation that only requires a GCC compiler (version 4.8 or higher) and OpenMPI version 1.8 or higher. While the code is not currently programmed to run in parallel, it has been set up such that is compiled with MPI so that future improvements in this area can be made. A GNU Makefile is provided in the `/src` directory of the package and can be executed in a Linux environment by simply typing `make` in the directory. Once this is done, an executable called `em3` will be produced. Simply execute `em3` in any directory containing an INPUT and CONFIG file to perform a calculation based on the settings in those files. The viability of running EM3 in a Windows or Mac environment is uncertain, but it's simply a C++ program with no external packages, so any compiler or operating system should work. The provided Makefile in the `/src` directory should be used in a Linux environment, however.

All the examples covered in this report are in the `/examples` directory. Each folder contains an INPUT and CONFIG file. After compiling EM3 and creating an `em3` executable, simply executing the executable in these directories will perform the MD simulation. For example, to simulate solid crystalline argon (c-Ar) at a density of $n = 1.09$ (in reduced LJ units) at 50 K, navigate to the `/examples/solid/n=1.09,T=50K` directory and execute `em3` in this directory. These examples and their results will be the subject of the next chapter.

# CHAPTER 2.  ARGON CALCULATIONS

## 2.1  Preliminary Checks

A classical test on the efficacy of a Verlet integrator and newly coded potential is checking energy conservation in the NVE ensemble. A model system to check energy conservation is a low temperature crystalline argon (c-Ar) system at 3 K. Argon is a face-centered cubic (FCC) lattice when solid, with an equilibrium lattice parameter of 5.24 Å, so this structure will be used for the low-temperature system. This simulation, and the following simulations in this chapter, will be equilibrated in a NVT ensemble with the Andersen thermostat of Equation 8 for 10 ps and then sampled for properties during a NVE production phase of 100 ps. The timestep will be 1 fs. The EM3 settings for this low temperature system is given by Figure 4

```
nsteps: 110000
nequil: 10000
nout: 100
max_neigh: 200
rc: 3
neighcount: 0
newton: 0
offset: 1
mass: 6.6335209e-26
temperature: 3.0
timestep: 1
epsilon: 1.65e-21
sigma: 3.4e-10
```

**Figure 4 – EM3 INPUT script settings for FCC argon at 3 K.**

Note that the temperature input into the EM3 program is in Kelvin units, so the input of 3 K for temperature will be converted to $T^* = \frac{3}{120} = 0.025$ in LJ units within the program.

The cutoff used through all simulations in this report is $r_c = 3\sigma$ (as seen in Figure 4), which was chosen to be longer than the typically recommended[7,9] range of $r_c = 2.5\sigma$. The structure of the system corresponds to the lattice constant of solid argon, which is 5.24 Å [10]. For a FCC unit cell with 4 atoms per cell, this yields a number density of $n = {}^4/_{(5.24\text{Å})^3} = 2.78 \times 10^{28} \frac{atoms}{m^3}$. In reduced LJ units this a density of $n = 1.09$. For all purposes of verifying the code in this chapter, a 500 atom system will be used with varying densities. The structure is generated by adding linear combinations of the FCC lattice vectors to fill in a $5 \times 5 \times 5$ unit cell, where each $1 \times 1 \times 1$ cell corresponds to a 4-atom FCC cell with a lattice constant corresponding to the desired density. This procedure will always yield a 500 atom system, but the density can be changed. For all the simulations in this report, the initial structure is a 500 atom FCC system with varying densities. For the $n = 1.09$ case, this 500 atom structure is shown in Figure 5, viewed using VMD[11].

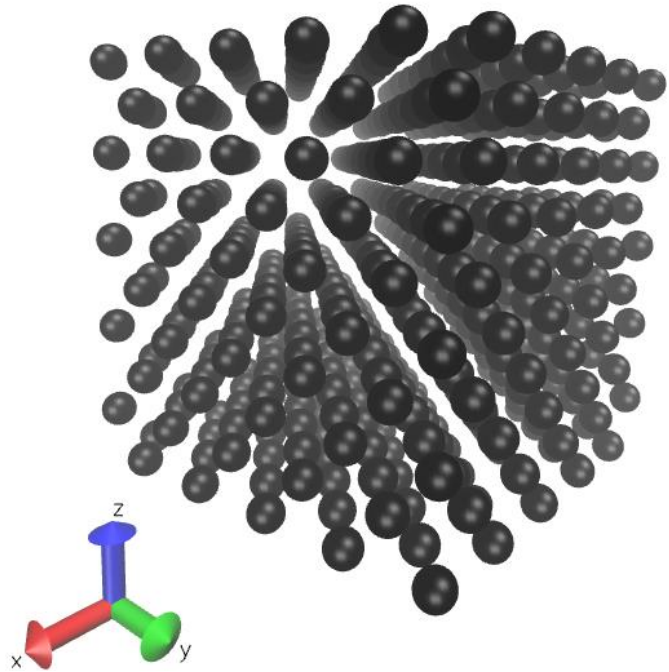

**Figure 5 – 500 atom FCC structure. This same structure and number of atoms are used throughout this report, except the density will change for different scenarios.**

Using this structure as the initial condition, a MD simulation was performed with 10 ps of NVT equilibration followed by 100 ps of NVE dynamics. A plot of the potential energy per atom, and total energy per atom is shown in Figure 6.

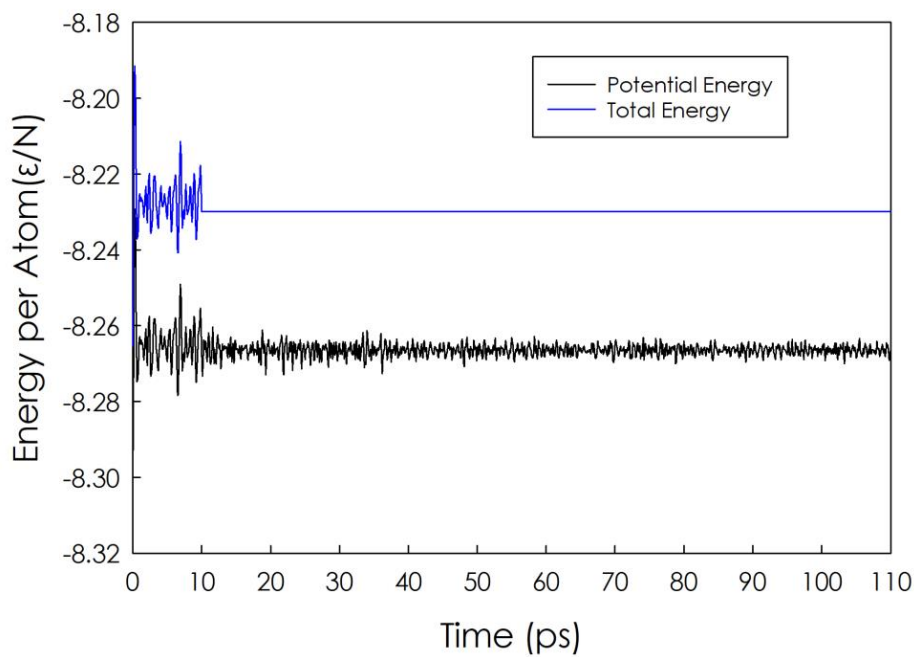

**Figure 6 – Potential and total energy (kinetic plus potential) for c-Ar at 3 K. The first 10 ps involved MD in the NVT ensemble with the Andersen thermostat, while the remainder of the simulation (10 ps to 110 ps) was NVE.**

The kinetic energy was excluded from Figure 6 since the difference in scales distracts from important information that is seen in the figure; the total energy is not conserved during the NVT portion (first 10 ps) while the total energy is conserved during the NVE portion (10 ps to 110 ps). It is important to note here that an offset was subtracted from each pair contribution to the total potential energy in Equation 2, which scales the LJ potential so that it is zero at the cutoff. This enforces energy conservation by eliminating the discontinuity that exists between the cutoff and value of zero for the potential, and is achieved by subtracting

$$U_{offset} = 4\left(\frac{1}{r_c^{12}} - \frac{1}{r_c^6}\right) \tag{14}$$

from every pairwise contribution to the total potential energy in Equation 11. This subtracting is achieved by setting the `offset` tag to zero in the input script, and a value of 1 will ignore the offset.

The efficacy of the thermostat and its effect on kinetic energy and temperature for the c-Ar system at 3 K is shown in Figure 7.

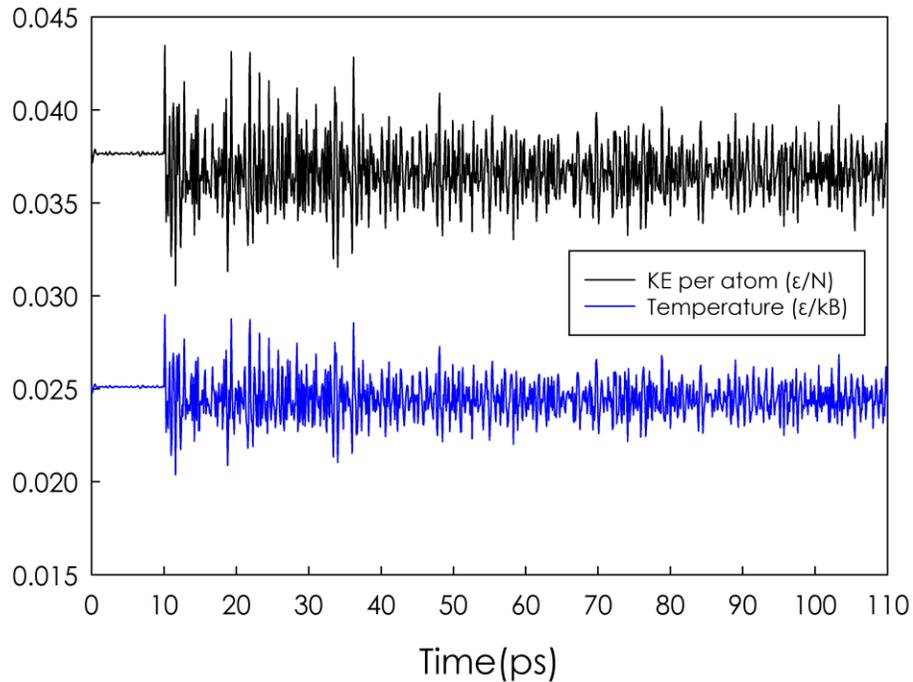

**Figure 7 – Kinetic energy and temperature in LJ units versus time for c-Ar at 3 K. This plot shows the effectiveness of the thermostat, which is turned on for the first 10 ps.**

Note that according to Equation 9 for kinetic energy and Equation 10 for temperature, the kinetic energy per atom is simply $\frac{3}{2}T$ in reduced LJ units, as seen in Figure 7. Figure 6 and Figure 7 illustrate the effectiveness, energy conservation, and thermostat efficacy of a simple low-temperature c-Ar system.

## 2.2   Solid-State Simulations

To further verify the proper dynamics of low temperature c-Ar, we can perform more simulations on the n = 1.09 system below the melting point of 58 K [12]. The temperatures of choice will be 10 K, 30 K and 50 K. Qualitatively, we should expect the kinetic energy and therefore total energy to rise with each case. Higher temperatures should also yield more variations in the potential energy. This is shown in Figure 8 for c-Ar at 10 K, 30 K and 50 K.

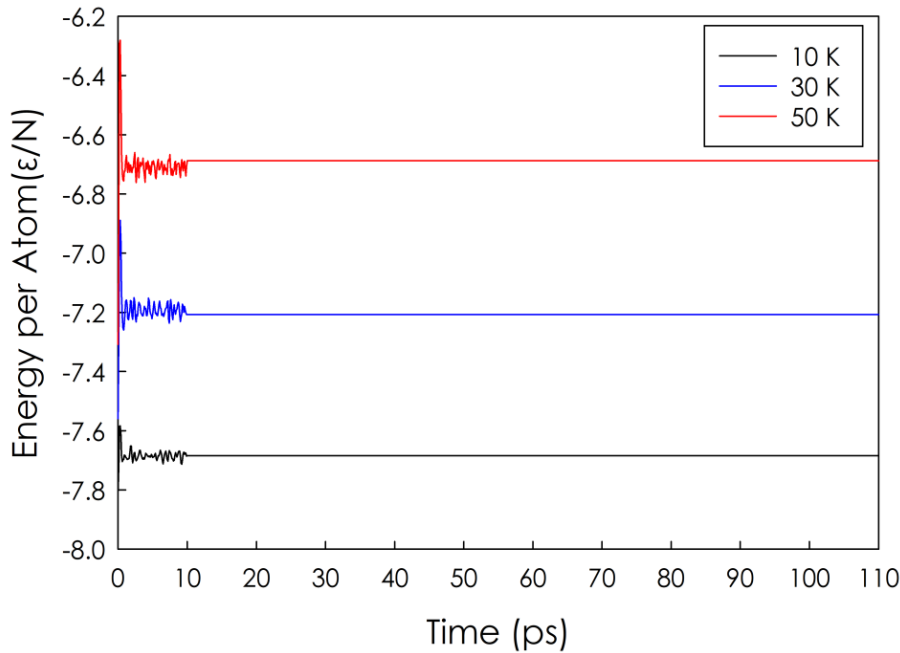

**Figure 8 – Total energy per atom (LJ units) versus time for c-Ar at 10 K, 30 K, and 50 K. The first 10 ps is NVT dynamics and NVE dynamics ensues for 10 ps to 110 ps.**

As shown in Figure 8, higher temperatures yield a higher total energy. More importantly, Figure 8 shows that total energy is still conserved for higher temperature c-Ar systems. Further qualitative checks can be realized by checking the potential energy as a function of time, since larger vibrations due to higher temperature should result in atoms moving at higher points in the LJ potential energy well. Higher temperatures for a crystal, on average, should therefore yield higher system potential energies which also fluctuate more. The potential energy versus time for c-Ar at 10 K, 30 K and 50 K is shown in Figure 9.

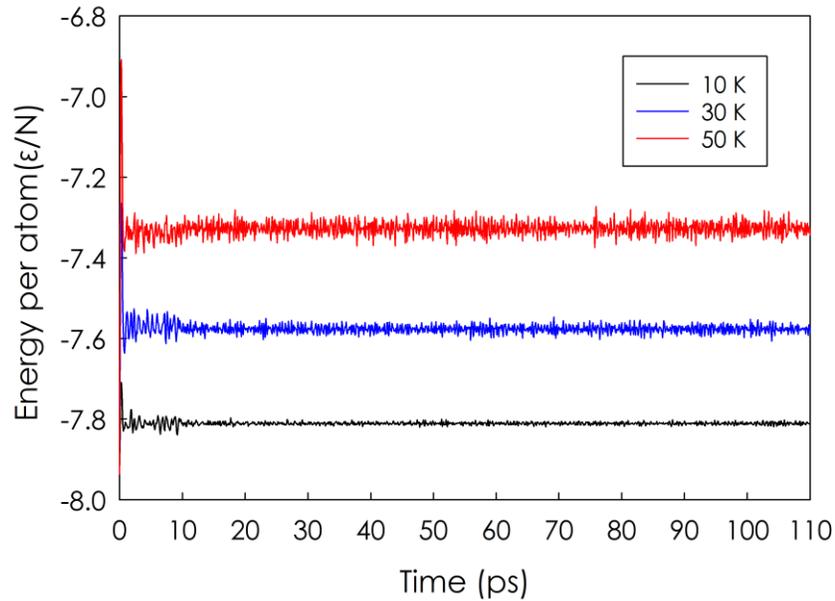

**Figure 9 – Potential energy per atom (LJ units) as a function of time for c-Ar at 10 K, 30 K and 50 K.**

Further insight into the reason behind higher potential energy for higher temperatures can be seen by observing the MSD given by Equation 12 as a function of time. Also as another qualitative check on the dynamics, the MSD for a crystal should fluctuate about some value since the system is remaining in its solid state. The MSD as a function of time is for c-Ar at 10 K, 30 K and 50 K is shown in Figure 10.

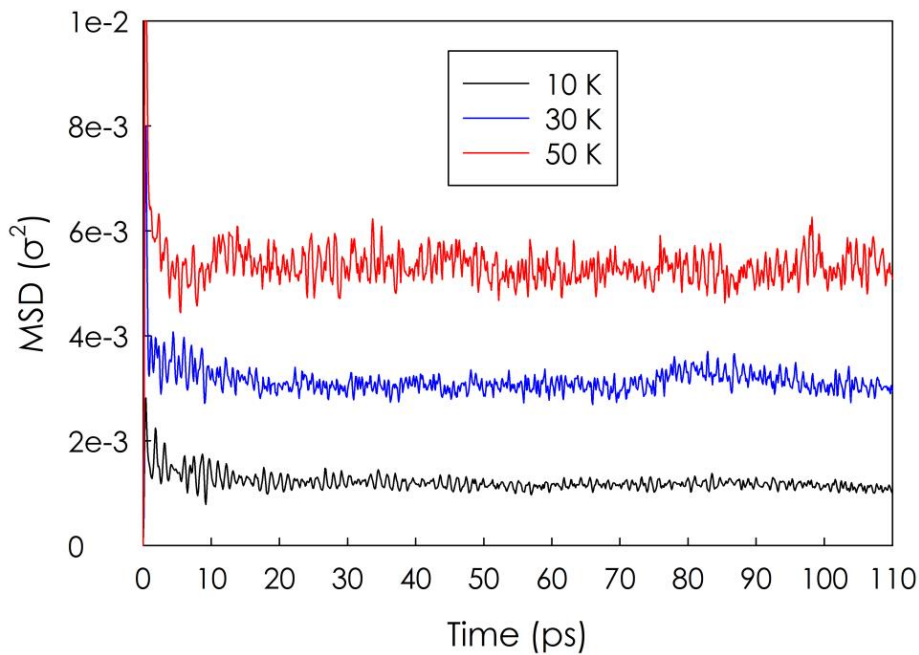

**Figure 10 – MSD (in LJ units) as a function of time for c-Ar at 10 K, 30 K, and 50 K.**

As expected, Figure 10 shows that atoms in the c-Ar crystal are vibrating further distances at higher temperatures. This explains the higher potential energy per atom from Figure 9. Figure 10 also further validates the EM3 code, which results in stable MD for a crystalline solid. Further qualitative checks on the thermostat of these higher temperature crystals is shown in Figure 11.

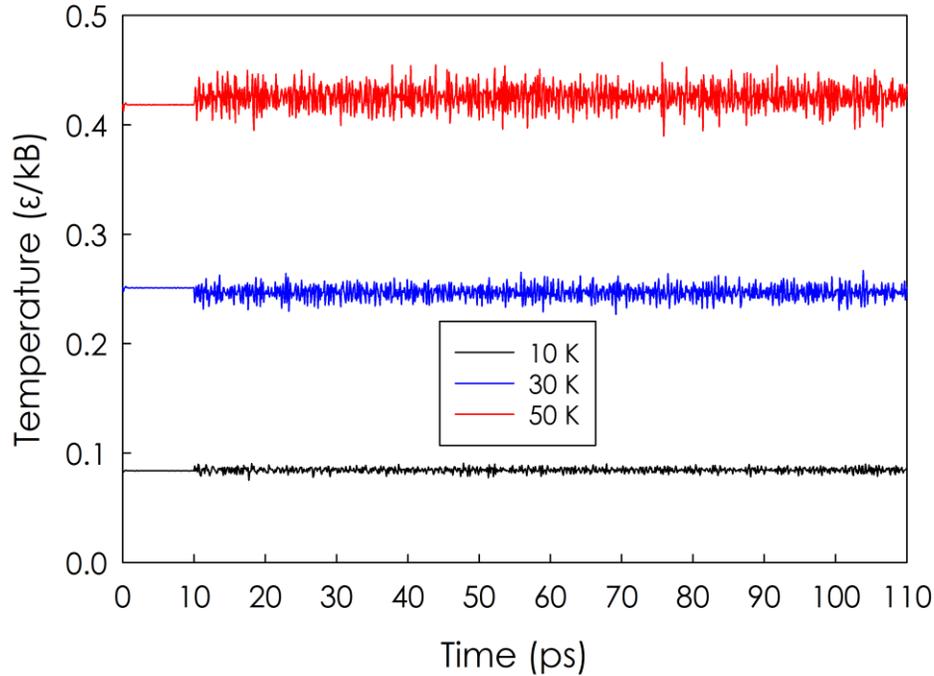

**Figure 11 – Temperature (in LJ units) versus time for c-Ar at 10 K, 30 K and 50 K. The first 10 ps is NVT dynamics and NVE dynamics ensues for 10 ps to 110 ps, as shown by the temperature trend.**

Figure 11 shows the efficacy of the thermostat for the first 10 ps NVT equilibration stage using the Andersen thermostat from Equation 8, followed by NVE dynamics where the temperature experiences larger fluctuations but still maintains a value about the desired starting temperatures. Note that the temperature in LJ units is scaled by $\varepsilon/k_B$, so that the values seen in Figure 11 are the values in K divided by 120 K. For example, a temperature of 30 K in LJ units is $30/120 = 0.25$, which agrees with the values in Figure 11 for the 30 K simulation.

To illustrate energy conservation in the code along, all the results up to now have included the offset subtracted from each pair contribution to the total potential energy.

When comparing with reference values for energies, however, it is useful and harmless to the dynamics to relieve this cutoff by setting the `offset` tag to 1 in the INPUT script. Relieving the offset and performing a static calculation (0 K) on a n = 1.09 FCC argon crystal with a cutoff of $r_c = 3\sigma$, the resulting potential energy per atom is -8.303 for the 500 atom system in Figure 5. The physical significance of this value for an empirical analytical interatomic potential is that the potential energy corresponds to the system cohesive energy – the energy required to separate a system of $N$ atoms completely into its individual constituents [13], i.e. $E_{coh} = E - NE_{atom}$, where $E_{atom}$ is the energy of an atom isolated by itself with no interactions ($E_{atom} = 0$ for a pair potential with no neighbors to interact with). Since analytical pair potentials do not say anything about the individual atom energies when they are not interacting with other atoms, the potential energy is simply the cohesive energy. Note that the cohesive energy is the energy *required to separate* a system of atoms, so we can simply flip the sign of the potential energy calculated from an empirical potential to get the cohesive energy. In this case for c-Ar at 0 K, the cohesive energy per atom is calculated to be 8.303 by the EM3 code (simply negative of the potential energy per atom). The calculated result at 0 K can therefore be compared with experimental values in literature for the cohesive energy. This result compared to experimental values from experiments[14] and *ab initio* calculations[15] are shown in Table 2.

**Table 2 – Comparison of cohesive energies per atom (meV/atom) between this work, experiment, and *ab initio* calculations from literature.**

|  | This work | Experiment[14] | *Ab initio*[15] |
|---|---|---|---|
| Cohesive energy per atom (meV) | 85.51 | 88.9 | 82.81 |

The value of 8.303 (in LJ units) calculated in this report at 0 K has been converted to meV in Table 2. The agreement with both experiment and *ab initio* are well within the range of experimental and *ab initio* agreements with each other, thus suggesting that the LJ potential is indeed coded correctly in the potential class.

With the efficacy of the EM3 code presented for low-temperature c-Ar, showing stable dynamics, energy conservation, and thermostat effectiveness, and proper cohesive energy calculation, it is now safe to move on to more complicated systems such as fluids. For more prudent calculations, we can compare with MC simulations of higher temperature argon systems at varying temperatures and densities.

## 2.3 Fluid Simulations

To more strictly test the accuracy of the code, it is worthwhile to compare to previously existing LJ calculations via different means, e.g. MC calculations and literature [16]. MC calculations are performed using the MC Fortran code from Allen et. al [9] and average quantities are taken over 1000 timesteps. We will use the three different cases for density and temperature described by Johnson et. al [16], so that we can also compare results with these. These three cases are (in LJ units): (1) a density n = 0.5 and temperature T = 5, (2) n = 0.9 and T = 2, and (3) n = 0.8 and T = 4. MD simulations for these three cases follow the same procedure described in Section 2.1, with a cutoff $r_c = 3\sigma$, 500 atoms, and the initial structure is an FCC lattice with a volume such that the proper density is achieved. To compare the average potential energy per atom to MC calculations and Johnson et. al [16], there will be no offset subtracted from the potential energy during these simulations (the `offset` tag is set to 1 in the INPUT scripts). Total energy will therefore not be conserved, but this will be due to atoms moving inside and outside of the cutoff radius and not the result of improper dynamics. The dynamics will therefore not be affected by ignoring the potential energy offset, and energy conservation of the LJ potential in the EM3 code has been readily demonstrated in Section 2.1. The simulations are first equilibrated in the NVT ensemble for 10 ps, using the velocity rescaling of Equation 8, followed by a 100 ps NVE run where average energies and pressures will be sampled. The timestep is set to 1 fs, and data is output every 100 timesteps.

As a first qualitative check, we expect the system to stray from its crystalline state due to the lower densities and higher temperatures of the three cases. This is readily seen by plotting the MSD versus time, as shown in Figure 12.

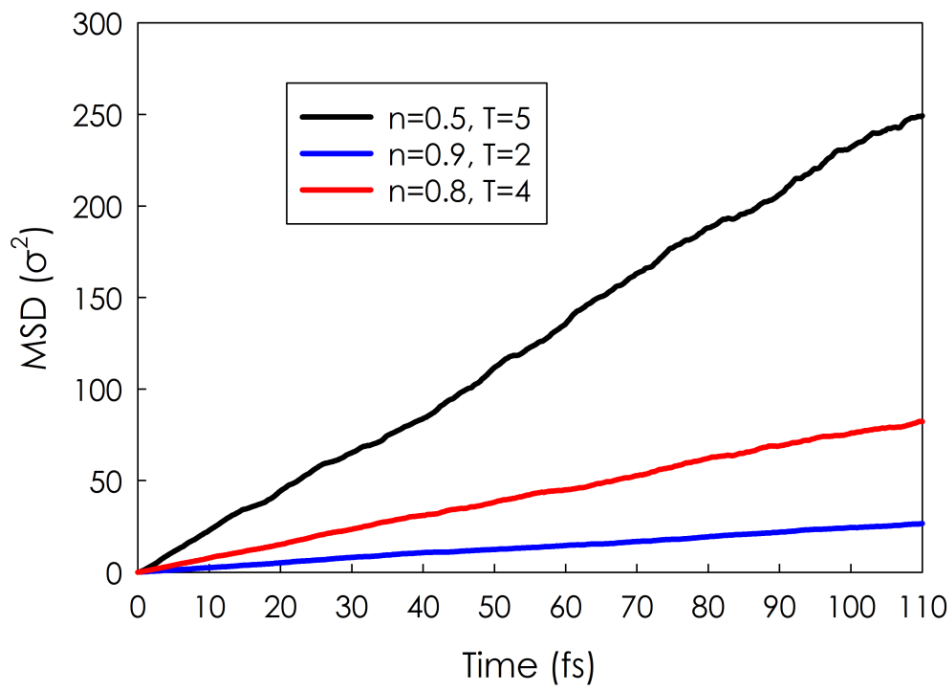

**Figure 12 – MSD (in LJ units) versus time in a MD simulation for the three cases of liquid argon with varying densities and temperatures.**

Unlike the MSD as a function of time for c-Ar, the MSD in these liquid states grows with time as shown in Figure 12. Figure 12 is therefore a sanity check that the system is indeed experiencing diffusive behavior as we would expect for temperatures above the melting point and densities below n = 1.09 for argon. It is also important to check the temperature of the simulation before we sample any relevant properties and claim that they are associated with a specific temperature, so the temperature during the simulation is plotted in Figure 13.

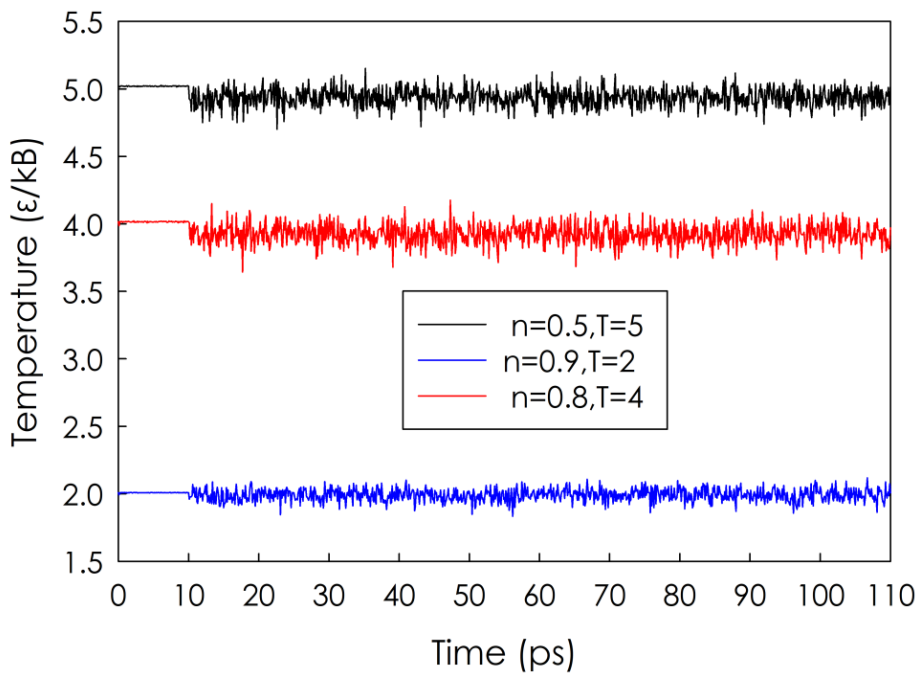

**Figure 13 – Temperature (in LJ units) as a function of time in a MD simulation for three cases of liquid argon with different densities and temperatures.**

Figure 13 shows that after the equilibration period during the first 10 ps, the NVE dynamics result in a stable temperature about the desired value for the remainder of the simulation. We can therefore claim that any properties sampled during the NVE portion of the simulation (10 ps to 110 ps) are associated with the desired temperature.

The first quantity to check will be potential energy per atom, which is shown in Figure 14 for all three liquid argon scenarios as a function of time.

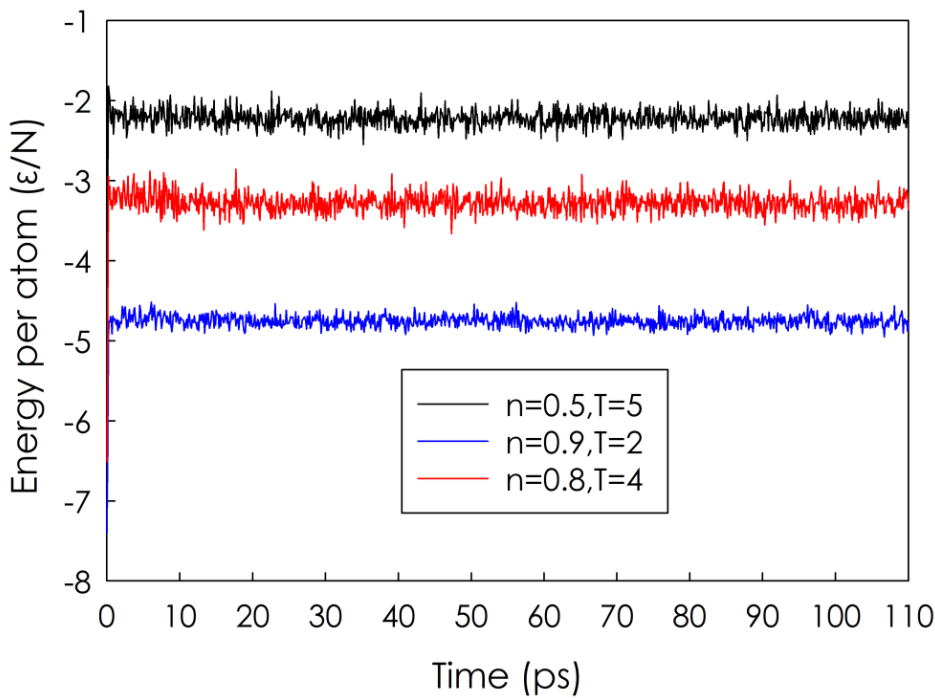

**Figure 14 – Potential energy per atom (in LJ units) as a function of time in a MD simulation for three cases of liquid argon with different densities and temperatures.**

The potential energy is shown to be well equilibrated at 10 ps and remains stable throughout the course of the simulation for all three cases. The average for each case during the NVE portion of the simulation (10 ps to 110 ps) is given in Table 3, along with literature[16] values and MC calculations.

**Table 3 – Average potential energy per atom (LJ units) for three liquid argon cases of varying density and temperature according to this report, MC calculations and literature.**

|  | This work (MD) | MC calculations | Johnson et. al[16] |
|---|---|---|---|
| $n = 0.5, T = 5$ | -2.22 | -2.36 | -2.4 |
| $n = 0.9, T = 2$ | -4.76 | -5.03 | -5.0 |
| $n = 0.8, T = 4$ | -3.28 | -3.5 | -3.5 |

The errors compared to literature values are 7.5%, 4.8%, and 6.8% for Cases 1, 2 and 3, respectively. While this validates the potential energy sampled in for three different liquid argon scenarios to literature values within 10%, the forces or derivatives of the potential can be validated by comparing the pressures sampled during these simulations, since pressure depends directly on the forces per Equation 4 and Equation 11. The pressure as a function of time is plotted for all three liquid argon scenarios in Figure 15.

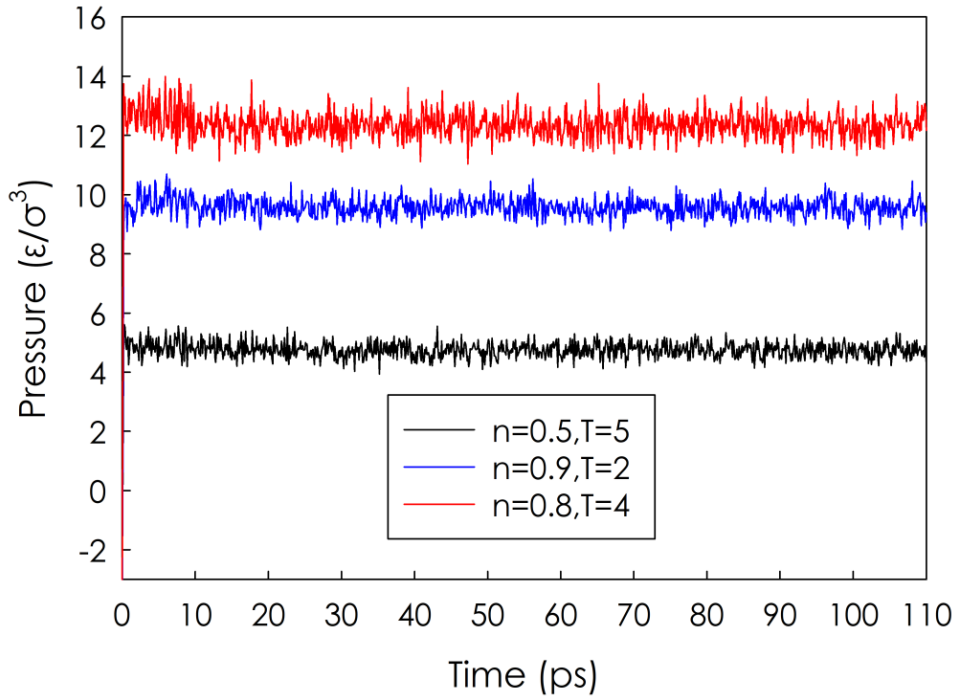

**Figure 15 – Pressure (in LJ units) versus time in a MD simulation for three different liquid argon scenarios.**

Like the potential energy calculations in Figure 14, the pressure is shown by Figure 15 to exhibit stable oscillatory behavior, thus exemplifying the stability of forces and dynamics in the liquid simulations. The average pressure past the NVT equilibration point (10 ps) in the NVE ensemble (10 ps to 110 ps) is calculated and compared to MC calculations and literature values from Johnson et. al[16] in Table 4.

**Table 4 – Average pressure (LJ units) comparisons from this report, MC calculations literature MC calculations.**

|  | This work (MD) | MC calculations | Johnson et. al[16] |
|---|---|---|---|
| $n = 0.5, T = 5$ | 4.74 | 4.67 | 4.7 |
| $n = 0.9, T = 2$ | 9.56 | 9.09 | 9.1 |
| $n = 0.8, T = 4$ | 12.32 | 12.1 | 12.1 |

The agreements errors between MD averages calculated in this report and MC values from literature are -0.8%, 5.1%, and 1.8% for Cases 1, 2, and 3, respectively. Long-range corrections could have been added to the MD results from Table 3 and Table 4 but the agreement of literature and MC values within 10%, along with energy conserving dynamics and a working thermostat, seems sufficient to validate the code. The decent agreement with MC and literature values stems from using a longer-than-normal cutoff of $r_c = 3\sigma$, thus providing less of a need for long-range energy and pressure corrections. With the potential energy and pressure verified against literature MC calculations within a maximum of 7.5% error according to Table 3 and Table 4, we can confidently use the EM3 code to simulate more unknown scenarios regarding argon. Given this assurance, we can now move forward to calculate a more unknown quantity with no reference to abide by.

## 2.4 Diffusion Coefficient

The goal here is to calculate the diffusion coefficient of argon at 158 K with a number density of $8 \times 10^{27}/m^3$, or 0.314 in LJ units. The same input script from Figure 4 is used, except the temperature is now set to be 158 K, which the EM3 code internally converts to a reduce LJ temperature of 1.32. The same settings are applied towards equilibration and production runs; a NVT ensemble equilibrates the system for 10 ps followed by NVE integration for 100 ps. The sampling of MSD, via Equation 12, for diffusion coefficient calculation occurs during the NVE portion (the production run). To first verify energy conserving dynamics, the total energy for the simulation is plotted in Figure 16.

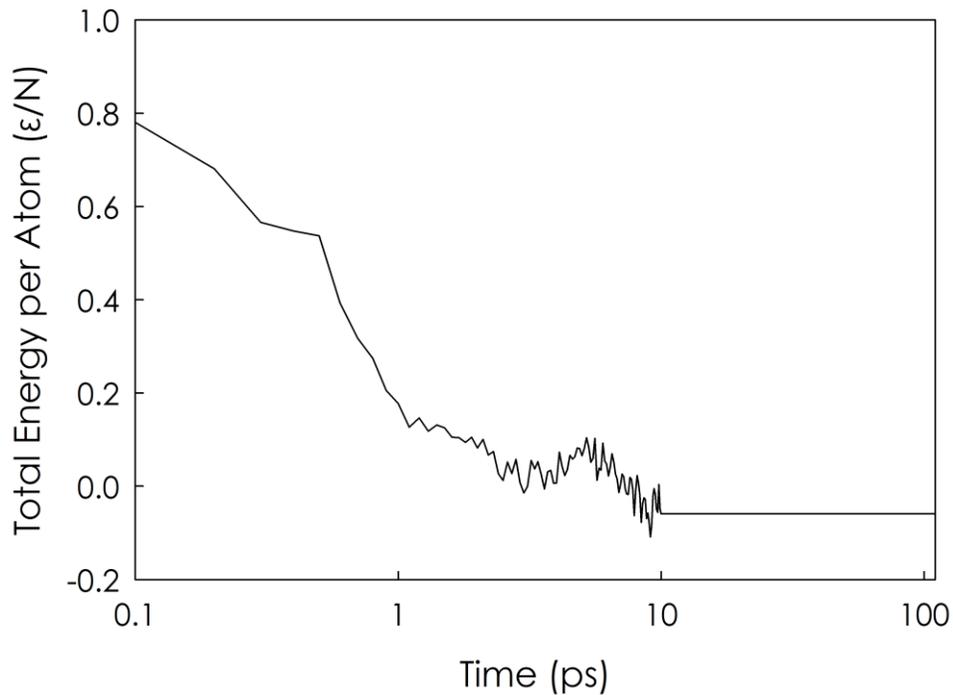

**Figure 16 – Total energy per atom (in LJ units) versus time for the 158 K argon system. The time axis is log-scaled to better show convergence in the equilibration regime (0-10 ps).**

It is evident through Figure 16, however, that this system was not finished equilibrating in total energy during the 10 ps NVT equilibration stage. This is in contrast to the other liquid argon systems mentioned in Section 2.3 most likely due to the lower density n = 0.314 (in LJ units) of this system. To more properly equilibrate the system, a longer equilibration should ensue. The simulation time settings were therefore changed to have a 100 ps NVT equilibration stage followed by a 1000 ps NVE run, where the MSD will be sampled to calculate the diffusion coefficient. The total simulation time is therefore 1100 ps. This simulation took 1 hour and 10 minutes using EM3. The total energy for these settings are shown in Figure 17.

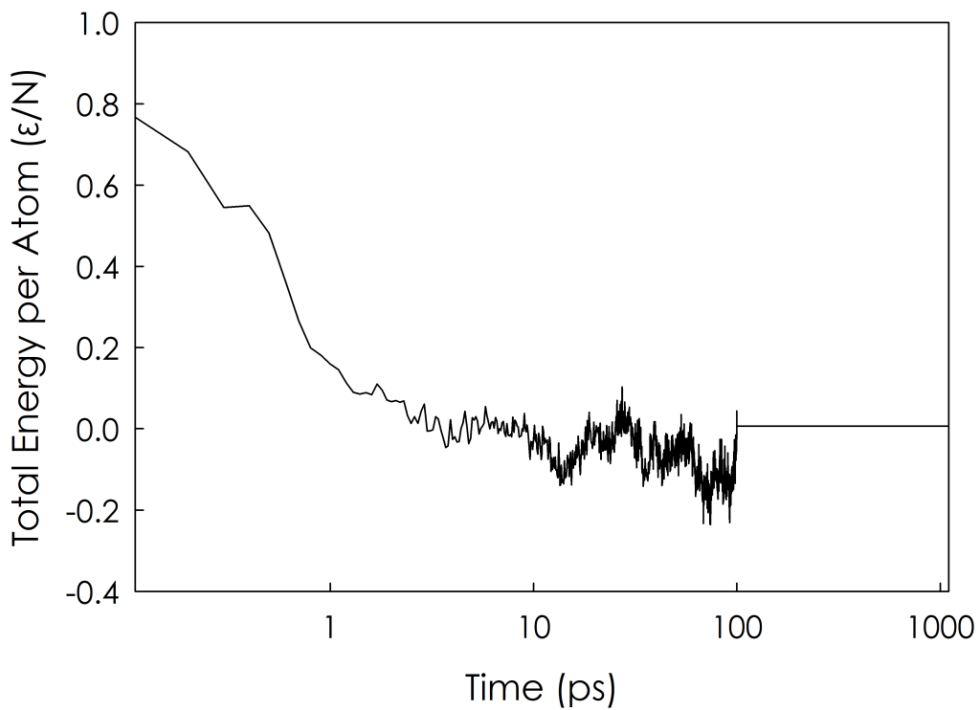

**Figure 17 - Total energy per atom (in LJ units) versus time for the 158 K argon system. The time axis is log-scaled to better show convergence in the equilibration regime (0-100 ps).**

Comparing Figure 16 and Figure 17, it is evident that the longer equilibration time necessary to further converge the total energy. As a further sanity check, total energy is seen to be conserved in the NVE portion of the simulation (100-1100 ps) in Figure 17. To ensure that we are sampling MSD at the proper temperature and that no drifts in temperature or kinetic energy are occurring, the temperature versus time for this simulation is shown in Figure X.

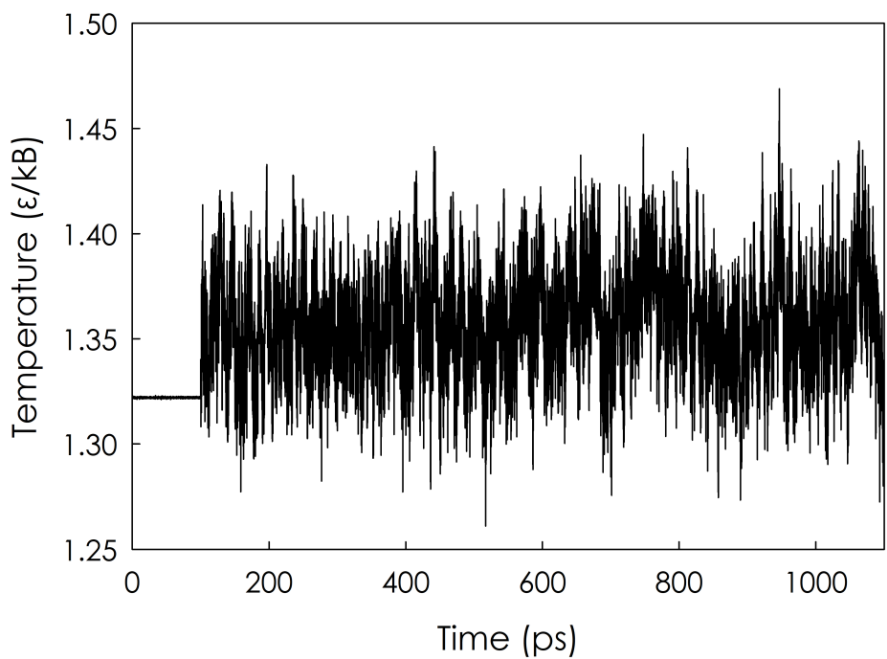

**Figure 18 – Temperature (in LJ units) versus time for argon at 158 K.**

Figure 18 shows that after the NVT equilibration stage, during the NVE run, the temperature is free to fluctuate near the desired value of 1.32 (in LJ units). Since the NVT equilibration successfully converged the total energy, total energy is conserved in the NVE run, and the NVE run contains temperatures fluctuating near the desired temperature, we can confidently use MSD data as a function of time to calculate the diffusion coefficient for this system. The MSD as a function of time is shown in Figure 19.

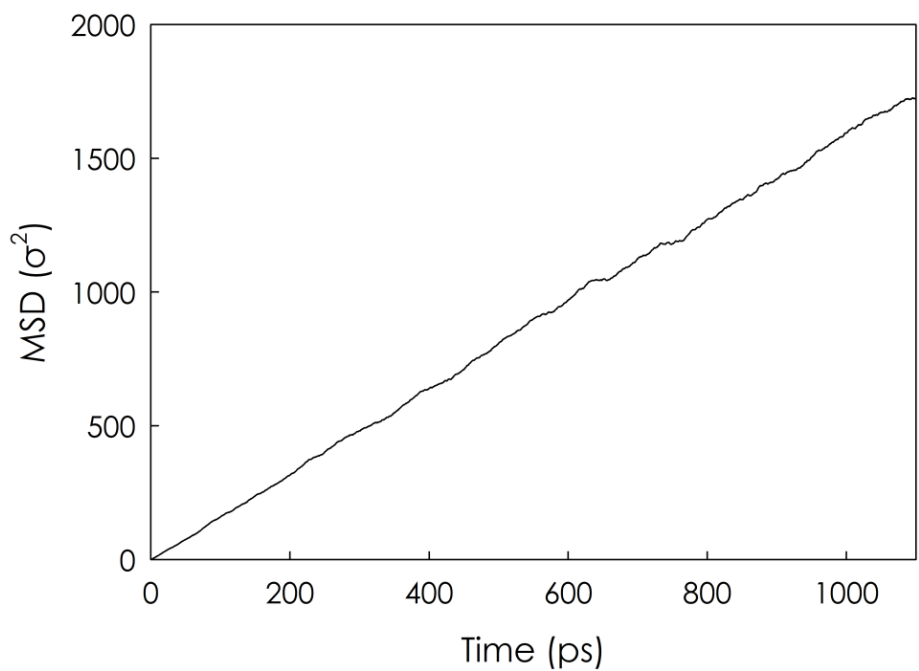

**Figure 19 – MSD (in LJ units) versus time for argon at 158 K. The data was sampled every 100 fs.**

The equilibration stage (0-100 ps) is not of use in the diffusion coefficient calculation, so we can fit a linear regression model to NVE stage (100-1100 ps) of the simulation. Linear regression is used to fit the data from the NVE portion of the simulation to a line. The resulting fitted line plotted against the data is shown in Figure 20.

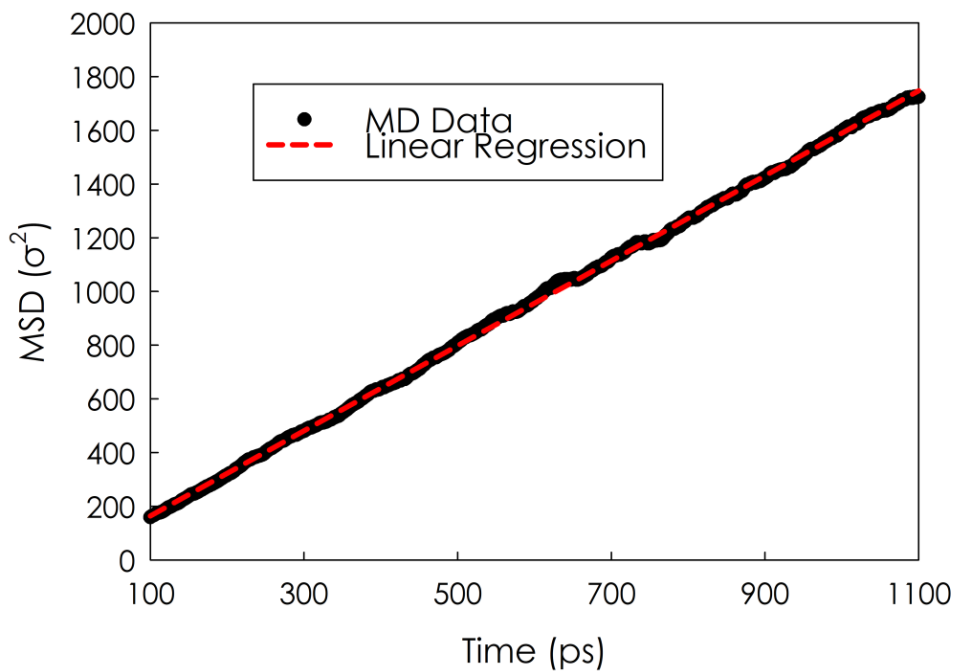

**Figure 20 – MSD (in LJ units) versus time during the NVE portion of the MD simulation (100-1100 ps). A linear regression fit to the data is shown as the dashed line.**

Results of the linear regression are shown in Table 5, where the coefficient of determination $R^2$ is shown to be close to unity and therefore a line excellent approximates the data.

**Table 5 – Linear regression parameters for fitting a line to MSD data as a function of time, along with the diffusion coefficient in different units.**

|  | Slope ($\sigma^2$/ps) | Intercept | $R^2$ | $D$ ($\sigma^2$/ps) | $D$ ($\sigma^2/\tau$) |
|---|---|---|---|---|---|
| Values | 1.5848 | 5.2225 | 0.9995 | 0.2641 | 5.7310 |

Now to apply the Einstein relation of Equation 13, we note that $\frac{\partial \langle \Delta r^2(t) \rangle}{\partial t}$ is the slope, so the diffusion coefficient is given by $D = {1.5848}/{6} = 0.2641$ $\sigma^2$/ps as shown in Table 5. To convert $\sigma^2$/ps purely to LJ units, we note that the LJ time unit $\tau = \sigma\sqrt{m/\varepsilon}$ from Table 1 has a value of $\tau = 2.17 \times 10^{-12}$ s for argon, which we can convert units to get $D = 5.7310$ $\sigma^2/\tau$ as shown in Table 5. We can qualitatively view this diffusion by calling the `output->write_xyz()` function in EM3, which will write XYZ formatted structure files that can be viewed in VMD[11]. The structures at different times for the whole 1.1 ns simulation are shown in Figure 21.

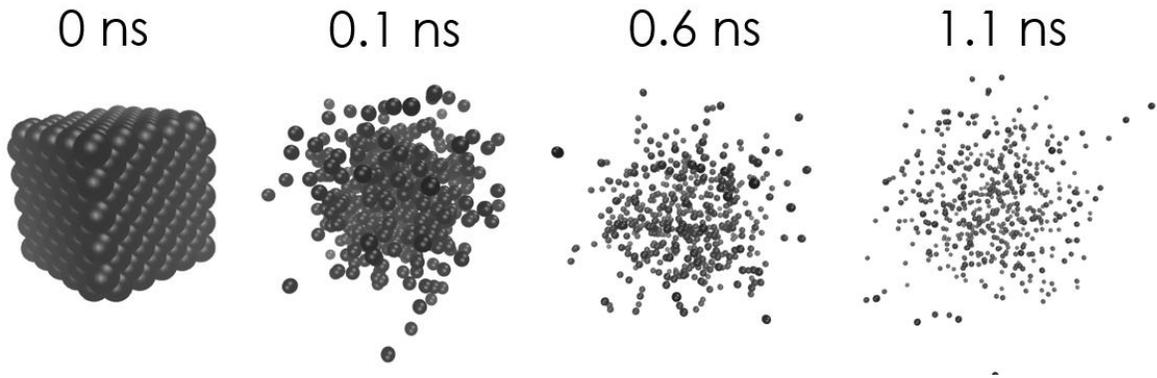

**Figure 21 – Visualization of the diffusion in liquid argon at 158 K with a density of 8 ×10²⁷/m³.**

The atom sphere sizes in Figure 21 are kept the same in all images, but the camera is zoomed out to view the diffusion. Using the minimum image convention from Equation 1, we can track atom positions in this manner without having to wrap them back around in the periodic box. This leads to the convenient calculation of the MSD as a function of time, and therefore the diffusion coefficient calculations in Table 5.

# REFERENCES

1       Plimpton, S. Fast parallel algorithms for short-range molecular dynamics. *Journal of computational physics* **117**, 1-19 (1995).

2       Lindahl, E., Hess, B. & Van Der Spoel, D. GROMACS 3.0: a package for molecular simulation and trajectory analysis. *Molecular modeling annual* **7**, 306-317 (2001).

3       Johnston, M. A., Galván, I. F. & Villà-Freixa, J. Framework-based design of a new all-purpose molecular simulation application: The Adun simulator. *Journal of computational chemistry* **26**, 1647-1659 (2005).

4       Hanwell, M. D. *et al.* Avogadro: an advanced semantic chemical editor, visualization, and analysis platform. *Journal of cheminformatics* **4**, 17 (2012).

5       Hutter, J., Iannuzzi, M., Schiffmann, F. & VandeVondele, J. CP2K: atomistic simulations of condensed matter systems. *Wiley Interdisciplinary Reviews: Computational Molecular Science* **4**, 15-25 (2014).

6       Einstein, A. Über die von der molekularkinetischen Theorie der Wärme geforderte Bewegung von in ruhenden Flüssigkeiten suspendierten Teilchen. *Annalen der physik* **322**, 549-560 (1905).

7       Frenkel, D. & Smit, B. *Understanding molecular simulation: from algorithms to applications*. Vol. 1 (Elsevier, 2001).

8       Tadano, T., Gohda, Y. & Tsuneyuki, S. Anharmonic force constants extracted from first-principles molecular dynamics: applications to heat transfer simulations. *Journal of Physics: Condensed Matter* **26**, 225402 (2014).

9       Allen, M. P. & Tildesley, D. J. *Computer simulation of liquids*. (Oxford university press, 2017).

10      Peterson, O., Batchelder, D. & Simmons, R. Measurements of X-ray lattice constant, thermal expansivity, and isothermal compressibility of argon crystals. *Physical Review* **150**, 703 (1966).

11      Humphrey, W., Dalke, A. & Schulten, K. VMD: visual molecular dynamics. *Journal of molecular graphics* **14**, 33-38 (1996).

12  Zha, C. S., Boehler, R., Young, D. & Ross, M. The argon melting curve to very high pressures. *The Journal of chemical physics* **85**, 1034-1036 (1986).

13  Kittel, C., McEuen, P. & McEuen, P. *Introduction to solid state physics*. Vol. 8 (Wiley New York, 1996).

14  Harl, J. & Kresse, G. Cohesive energy curves for noble gas solids calculated by adiabatic connection fluctuation-dissipation theory. *Physical Review B* **77**, 045136 (2008).

15  Rościszewski, K., Paulus, B., Fulde, P. & Stoll, H. Ab initio calculation of ground-state properties of rare-gas crystals. *Physical Review B* **60**, 7905 (1999).

16  Johnson, J. K., Zollweg, J. A. & Gubbins, K. E. The Lennard-Jones equation of state revisited. *Molecular Physics* **78**, 591-618 (1993).